# Unraveling cell-cell communication with NicheNet by inferring active ligands from transcriptomics data

Chananchida Sang-aram[1,2], Robin Browaeys[2,3], Ruth Seurinck[1,2], Yvan Saeys[1,2]

[1] Data mining and Modelling for Biomedicine, VIB Center for Inflammation Research, Ghent, Belgium
[2] Department of Applied Mathematics, Computer Science and Statistics, Ghent University, Ghent, Belgium
[3] BioIT Expertise Unit, VIB Center for Inflammation Research, Ghent, Belgium

## Abstract

Ligand-receptor interactions constitute a fundamental mechanism of cell-cell communication and signaling. NicheNet is a well-established computational tool that infers ligand-receptor interactions that potentially regulate gene expression changes in receiver cell populations. Whereas the original publication delves into the algorithm and validation, this paper describes a best practices workflow cultivated over four years of experience and user feedback. Starting from the input single-cell expression matrix, we describe a "sender-agnostic" approach which considers ligands from the entire microenvironment, and a "sender-focused" approach which only considers ligands from cell populations of interest. As output, users will obtain a list of prioritized ligands and their potential target genes, along with multiple visualizations. In NicheNet v2, we have updated the data sources and implemented a downstream procedure for prioritizing cell-type-specific ligand-receptor pairs. Although a standard NicheNet analysis takes less than 10 minutes to run, users often invest additional time in making decisions about the approach and parameters that best suit their biological question. This paper serves to aid in this decision-making process by describing the most appropriate workflow for common experimental designs like case-control and cell differentiation studies. Finally, in addition to the step-by-step description of the code, we also provide wrapper functions that enable the analysis to be run in one line of code, thus tailoring the workflow to users at all levels of computational proficiency.

# Introduction

Cell-cell communication (CCC) plays a pivotal role in all biological processes, making its study essential for a comprehensive understanding of complex biological systems. One mode of communication occurs when protein ligands secreted or expressed on the membrane of sender cells bind with receptors on receiver cells. These ligand-receptor interactions elicit a cascade of responses in the receiver cell, often altering changes in their gene expression levels. When using transcriptomics data to investigate CCC, it is therefore crucial to also consider this transcriptional response, as the true concentration of protein ligands and receptors may not be captured at the RNA level. We developed NicheNet[1] to fill this gap in computational tools for CCC inference, as preexisting tools only considered the expression levels of ligands and receptors. In contrast, NicheNet uses a prior model of ligand-to-target regulatory potential to prioritize which ligands are responsible for the observed transcriptional response in the receiver cell. To construct the prior model of NicheNet, we integrated three types of databases—ligand-receptor interactions, signaling pathways, and transcription factor (TF) regulation—to form a complete communication network spanning from ligands to their downstream target genes.

In this protocol, we describe each step of the NicheNet workflow in detail, enabling individuals without a bioinformatics background to conduct the analysis. We guide users through the basics of inferring top ligands and creating informative visualizations from their transcriptomics data. Furthermore, we explore additional features from NicheNet v2, including the prioritization of cell-type-specific ligand-receptor pairs and the integration of external data sources. NicheNet v2 is also accompanied by an update of data sources that has effectively doubled the number of ligands in the prior model compared to v1.

Overview of the procedure

NicheNet is a computational tool for studying intercellular communication. A typical workflow is shown in **Figure 1a**, where feature extraction is first performed (Steps 1-11), followed by two NicheNet analyses, ligand activity analysis and target gene prediction (Steps 12-15).

As input to NicheNet, users must provide cell type-annotated expression data that reflects a CCC event. As a result, most steady state data may not be suitable for a NicheNet analysis due to the inherent difficulty of distinguishing between cell-extrinsic and intrinsic signals (see **Figure 2**). The input can be single-cell or sorted bulk data from human or mouse. As output, NicheNet returns the ranking of ligands that best explain the CCC event of interest, as well as candidate target genes of these ligands. The subsequent paragraphs provide a detailed explanation of each intermediate step in the analysis.

After loading in required libraries and the input object (Step 1-3), users must select a "receiver" cell type whose gene expression is likely to be affected by the CCC event of interest (Step 4). In case of multiple receiver cell types, the workflow can be repeated for each receiver. First, **feature extraction** is performed to extract a list of (a) potential ligands, (b) a gene set that captures the downstream effects of the CCC event of interest, and (c) a background set of genes (**Figure 1b**). Both the data acquisition and research question will determine how these features are defined (**Figure 2**).

The initial list of potential ligands consists of ligands whose cognate receptors are expressed in the receiver cell type (Steps 5-7). By default, expressed genes are defined as genes with one or more transcripts in at least 10% of the cells from a cell type. To link specific ligands and receptors, we provide a ligand-receptor network that integrates interaction information from multiple databases. Although this initial *sender-agnostic* approach does not take the ligand expression into account, it can reveal active ligands with minimal or no RNA expression that may have been present at the protein level. Moreover, this approach will also consider ligands expressed by non-profiled cell types. However, if specific sender cell population(s) are known and captured in the dataset, users may also use the *sender-focused* approach, where ligands that are not expressed in the chosen sender cell type(s) are discarded (Step 8). When using the sender-focused approach, we recommend comparing ligand rankings to those from the sender-agnostic approach to assess how well a sender-specific ligand ranks in comparison to the global perspective (Step 24).

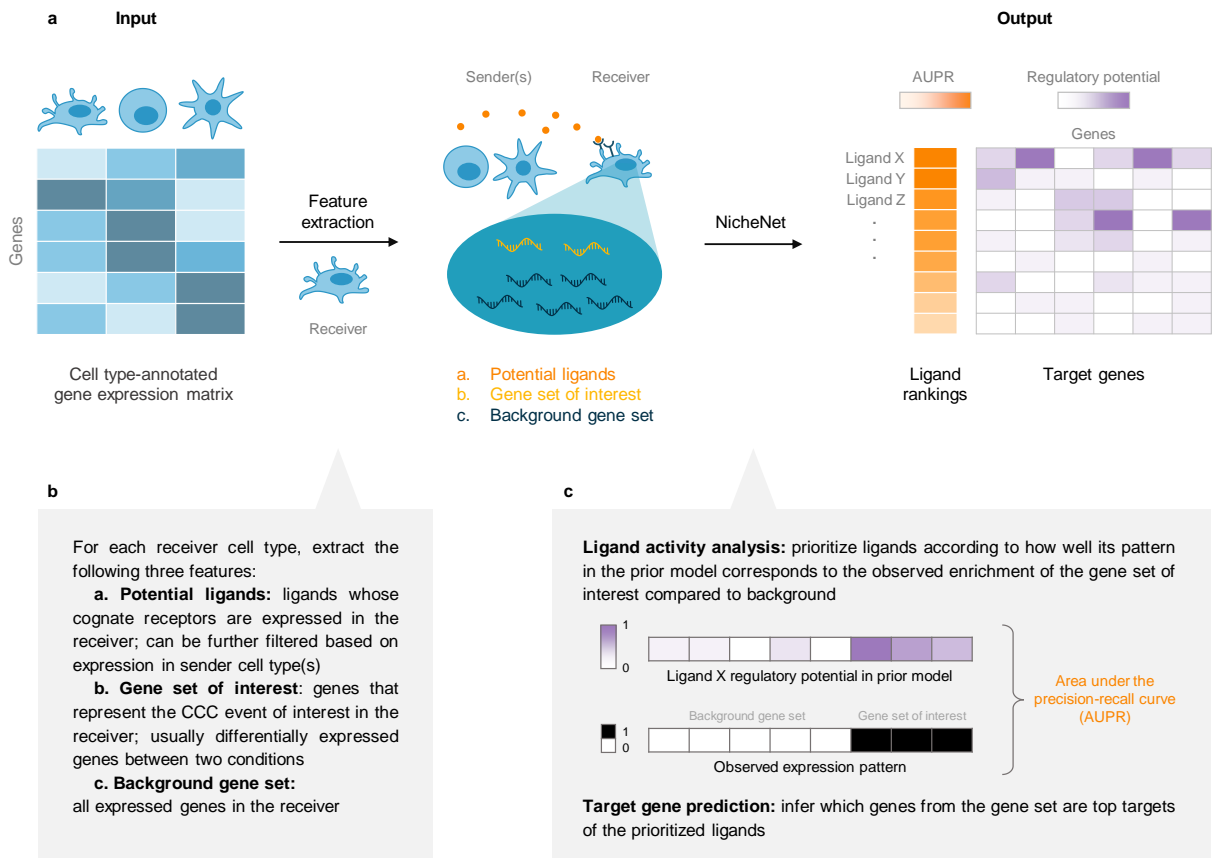

**Figure 1. Overview of a NicheNet analysis. (a)** Given an annotated gene expression matrix as input, NicheNet returns a ranking of ligands and their target genes responsible for the cell-cell communication response observed in the receiver cell population. **(b)** Feature extraction is performed to obtain a set of potential ligands, gene set of interest, and background gene set. **(c)** For each potential ligand, its ligand activity (i.e., area under the precision-recall curve) is calculated and used to determine its ranking.

Next, we define the gene set of interest as genes within the receiver cell type that are likely to be influenced by the CCC event (Steps 9-10). In typical case-control studies, we use the differentially expressed (DE) genes between the two conditions in the receiver cell type, assuming that the observed DE pattern is a result of the CCC event. Common experimental designs include data collected before and after ligand induction, infections, or even tumor development. In more complex designs, such as having multiple control and one treatment condition, users can instead perform a one-vs-all DE analysis. In datasets with only one condition (i.e., steady state), it is possible in certain biological scenarios to use the DE between two cell populations as the gene set of interest. For instance, if the user wishes to study a differentiation process that is influenced by the microenvironment, the DE genes between a progenitor and differentiated cell type could be used. Another scenario includes studying the differences between cell types that share a common ontogeny but are located in different niches, as with the subtypes of tissue-resident macrophages[2].

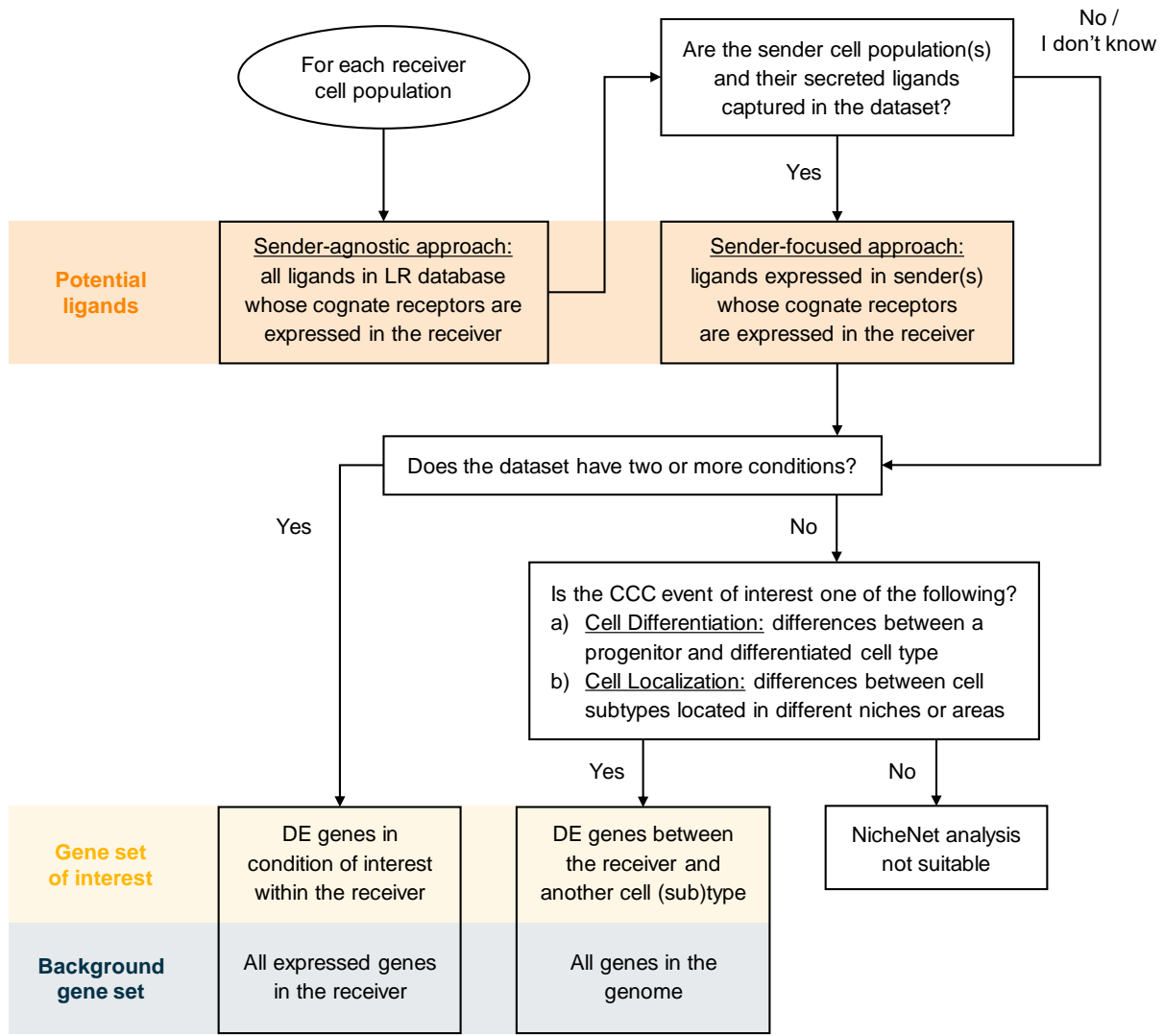

**Figure 2. Flowchart for feature extraction.** A guideline on extracting the potential ligands, gene set of interest, and background gene set. This also helps determine whether a NicheNet analysis is suitable for your dataset. LR: ligand-receptor; CCC: cell-cell communication; DE: differentially expressed.

Finally, the background gene set consists of genes unaffected by the CCC event (Step 11). This background set is comparable to the one in gene set enrichment analysis, where a common practice is to select genes that are detected at the level where they have a chance of being classified as differentially expressed[3]. In the case-control scenario, this corresponds to selecting all expressed genes in the receiver cell type. However, the size of the background set also impacts the analysis, as it should always be sufficiently larger than the gene set of interest. Thus, in the steady state case, where there is typically a substantial number of DE genes between two cell populations, we recommend users to instead use the entire genome as the background set.

Using these three features, the **ligand activity analysis** ranks the potential ligands based on how well the enrichment of the gene set of interest corresponds with the prior model of NicheNet (**Figure 1c**; Steps 12-14). To do so, we concatenate and binarize the gene set of interest and background gene set, where the "value" of a gene is set to 1 if it belongs to the gene set of interest and 0 otherwise. This binarization helps mitigate the noisiness of expression data, thus improving the robustness of the predictions. The resulting binary vector is referred to as the "observed expression pattern" (OEP). Meanwhile, the NicheNet prior model is a ligand-by-target matrix, containing continuous values that represent the supporting evidence that a specific ligand can regulate a given target gene. We subset the columns of this matrix to contain the same genes as the OEP. Then, for each ligand in the potential ligands list, we calculate the area under the precision-recall curve (AUPR) between the OEP and the relevant row from the ligand-target matrix. The AUPR is then corrected for random prediction by subtracting the proportion of positive classes (ones) in the OEP. This corrected AUPR value represents the **ligand activity** of the ligand. We selected the AUPR as the metric for ligand activity based on validation datasets (see Comparison with other methods). Note that for the v1 prior model, the Pearson correlation was instead determined as the best metric.

After obtaining the ligand rankings, the **target gene prediction** identifies genes that are the targets for each of the top 30-ranked ligands (Step 15). This is achieved by calculating the intersection between the gene set of interest and the 250 target genes with the highest regulatory potential in the prior model for the respective ligand. Users can adjust the number of genes if 250 would be too stringent. Additionally, users can retrieve a list of all the receptors associated with the top-ranked ligands (Step 16). We provide a variety of visualizations, such as chord diagrams and heatmaps, to present these and previous results (Steps 17-31, **Figure 3a-h**).

In v2, we introduce an additional downstream analysis to prioritize **cell type-specific ligand-receptor pairs** based on their expression levels (Steps 32-34). This analysis is derived from MultiNicheNet[4], the multi-sample multi-condition extension of NicheNet. In addition to the ligand activity calculation, which is based on the binarized expression pattern (**Figure 1c**), this prioritization approach also takes into account both cell type specificity and condition specificity of each ligand and receptor.

Cell type specificity is quantified by the upregulation and relative expression of a ligand or receptor in a particular cell type, in the condition of interest (if applicable). To calculate upregulation, we first perform a DE analysis between each cell type and the rest of the dataset (using the Wilcoxon signed-ranked test by default). The p-value is transformed by multiplying its negative log10 value with the sign of the fold change, thus giving the highest scores to strongly upregulated genes. Finally, rank-based scaling is applied to the transformed p-values across ligand-sender or receptor-receiver combinations. Since the DE analysis may be influenced by outlier cells with high expression, we also use relative expression as another criteria for cell type specificity. This involves averaging the log-expression values of each ligand/receptor, followed by rescaling (i.e., min-max normalization) across all cell types in the dataset.

Condition specificity is quantified by the upregulation of a ligand/receptor in the condition of interest compared to the reference condition. This is calculated by performing a DE analysis between two conditions across all cell types, followed by p-value transformation and rank-based scaling as described above. Note that since cell type specificity is only calculated in the condition of interest, there is no criterion that accounts for cell type-condition specificity.

In total, there are four criteria for cell type specificity (ligand upregulation, ligand expression, receptor upregulation, and receptor expression), two criteria for condition specificity (ligand and receptor), and one for ligand activity (after standardization and rescaling). A weighted aggregation of the criteria is performed to get a final prioritization score for each sender-receiver-ligand-receptor combination. For user convenience, we provide two scenarios with predetermined weights: a "case-control" scenario with all weights set to one and a "one condition" scenario where the condition specificity weights are set to zero. Users can customize the weight of each criterion, for example, prioritizing uniformly highly expressed genes by assigning a higher weight to the average expression criteria. Finally, we provide a specific visualization, called a mushroom plot, that incorporates the rankings and expression of each ligand-receptor pair (Step 35, **Figure 3i**).

Applications of the method

NicheNet has been applied in a wide range of biological studies—including the tumor microenvironment, cell type ontogeny, and post-injury recovery—with several of them experimentally validating NicheNet predictions[2,5–15]. For instance, Bonnardel et al. validated the importance of DLL-Notch signaling in monocyte-to-Kupffer cell (KC) differentiation in the liver by co-culturing monocytes with DLL4-expressing cells and adding BMP9 to the culture[5]. Furthermore, Guilliams et al. used KC-specific ALK1-knockout mice and an *in vivo* sequestration of BMP9/10 in stellate cells to demonstrate the crucial role of the ALK1-BMP9/10 axis in Kupffer cell maintenance and differentiation[2]. In a siRNA knockdown experiment, Karras et al. inactivated Notch3 in melanoma cells and found that their growth was no longer stimulated by endothelial cells[7].

Aside from being applied in wet lab studies, many computational tools also use NicheNet as part of their algorithm[16–22]. Scriabin[16] is a tool that infers CCC at single-cell resolution without cell aggregation or downsampling, using NicheNet in a crucial step to select ligands that can explain each cell's observed gene expression. Other tools like exFINDER[17] and LRLoop[18] instead integrate the NicheNet prior model with other data sources to identify external communication signals and predict feedback loops, respectively. As such, both the NicheNet ligand activity analysis and prior model serve as an invaluable tool for computational biologists.

Comparison with other methods

The most popular category of cell-cell communication inference tools are the so-called "expression permutation" tools[23]. These tools calculate the communication score between ligand-receptor (LR) pairs and evaluate its significance using permutation tests. The scoring methods can vary in complexity, from calculating the gene expression means as in CellPhoneDB[24], to mass action-based modeling in CellChat[25]. In addition to their popularity, these two tools consider multimeric receptors in their databases.

NicheNet can be considered a complement to expression permutation analysis. As NicheNet goes beyond ligand-receptor expression to also consider the target genes—that is, the potential downstream effects of these LR interactions—there is often limited overlap with results from expression permutation tools. Other tools that integrate intracellular signaling include CytoTalk[26], which constructs signal transduction networks *de novo* and identifies candidate pathways, scSeqComm[27], which in addition to scoring ligand-receptor pairs also quantifies intracellular signaling as a function of receptor-TF association and TF activity, and scMLnet[28], which uses the input single-cell data to generate a ligand-to-target matrix. In contrast to these tools, a major strength of NicheNet is that it has been systematically benchmarked using *in vitro* ligand treatment datasets, i.e., cell lines that have been treated with a ligand. NicheNet was able to predict which ligand has been added to the cell culture, as well as which target genes are being affected by this ligand. Furthermore, NicheNet has been biologically validated in many experiments, both *in vitro* and *in vivo*, that we further discuss in the next section.

Limitations

Steady-state data

We generally do not recommend applying NicheNet on steady-state data because a key assumption of NicheNet is that that the gene set of interest represents changes in the gene expression data of the receiver cell population caused by CCC from ligands in the microenvironment. Such changes are usually not captured in steady-state data, except in specific biological scenarios discussed in the paragraph on target gene selection.

Prior model

The ligand-target prior model of NicheNet is a matrix containing the regulatory potential, or the evidence of regulation between ligands and each gene in the transcriptome (including 1,287 ligands and 22,521 genes in the v2 model). During the construction of the prior model, no distinction was made on whether a ligand had an activating or inhibitory effect on its downstream targets. Therefore, both types of regulation additively contributed to the regulatory potential. As a result, users are unable to filter for a specific type of regulation when performing a ligand activity analysis. Furthermore, regulatory potential values can only be compared within a single ligand. That is, if the regulatory potential of ligandX-targetA > ligandY-targetA, it does not mean ligandX has greater potential than ligandY to regulate targetA; however, if ligandX-targetA > ligandX-targetB, it indicates that ligandX has greater potential to regulate targetA than targetB. Finally, users must take into account that the prior model only includes protein ligands that are secreted or form part of the cell membrane, and that the prior model lacks context specificity, as it was constructed across various cell types and conditions. However, users also have the flexibility to add or replace part of the prior model with their own data sources, since the prior model was constructed in a modular manner (Box 1). For instance, users could enhance the model's context specificity by integrating gene regulatory networks derived from their own epigenetics data.

## Materials

### Equipment

**Hardware**

Windows, Linux or MacOS

**Software**

- **R**: can be downloaded at https://cran.r-project.org/. The code in this paper was run with R version 4.3.2.
- **RStudio:** while not required, this is highly recommended for an easier working environment. RStudio Desktop can be downloaded at https://posit.co/products/open-source/rstudio/.
- **NicheNet** is implemented as an open-source R library that can be installed by running the following command in R:

```r
if (!requireNamespace("devtools", quietly=TRUE))
    install.packages("devtools")
devtools::install_github("saeyslab/nichenetr")
```

**Input data**

To follow the procedure below, users must provide a Seurat object (at least v3) containing the gene expression matrix with cell type annotations. However, this pipeline can be easily adapted to work with expression data stored in other formats (see Troubleshooting).

### Equipment Setup

**Example input data**

As example expression data of interacting cells, we will use mouse NICHE-seq data from Medaglia et al.[29] to explore intercellular communication in the T cell area in the inguinal lymph node before and 72 hours after lymphocytic choriomeningitis virus (LCMV) infection. Specifically, we will prioritize which ligands can best explain the downstream changes after LCMV infection in CD8 T cells as the receiver population. This dataset contains 13,541 genes and 5,027 cells from 6 cell populations: CD4 T cells (including regulatory T cells), CD8 T cells, B cells, NK cells, dendritic cells (DCs) and inflammatory monocytes.

The data can be downloaded by clicking "seuratObj.rds" at https://zenodo.org/record/3531889. To download the file with the command line, commands like `wget` or `curl` can be used (Linux/macOS):

```
wget https://zenodo.org/record/3531889/files/seuratObj.rds
```

To download the file within the R session, entering the following commands:

```r
options(timeout = 3600) # increase time limit for downloading the data
seuratObj <- readRDS(url("https://zenodo.org/record/3531889/files/seuratObj.rds"))
```

If the file is downloaded within the R session, it will have to be downloaded again once the session is restarted.

**Human/Mouse networks**

Three networks are required to run the NicheNet analysis: the ligand-target prior model, the ligand-receptor network, and the weighted ligand-receptor network. We provide these networks with either human or mouse gene symbols and they can be downloaded at

[https://zenodo.org/record/7074291/](https://zenodo.org/record/7074291/). This repository also contains other networks not used in the main analysis but in an additional visualization step (Step 30) and during model construction (Box 1). As with the example data, they can be downloaded locally through the website or command line, or in the R session. We recommend users to download them locally for convenience.

```
# Downloading data locally
ZENODO_PATH=https://zenodo.org/record/7074291/files
# Human ligand-target, ligand-receptor, and weighted networks
wget $ZENODO_PATH/ligand_target_matrix_nsga2r_final.rds
wget $ZENODO_PATH/lr_network_human_21122021.rds
wget $ZENODO_PATH/weighted_networks_nsga2r_final.rds
# Mouse ligand-target, ligand-receptor, and weighted networks
wget $ZENODO_PATH/ligand_target_matrix_nsga2r_final_mouse.rds
wget $ZENODO_PATH/lr_network_mouse_21122021.rds
wget $ZENODO_PATH/weighted_networks_nsga2r_final_mouse.rds

# Download the file in the R session, human example
# We assume these variables exist in the "Procedure" section
zenodo_path <- "https://zenodo.org/record/7074291/files/"
ligand_target_matrix <- readRDS(url(paste0(zenodo_path,
      "ligand_target_matrix_nsga2r_final_mouse.rds")))
lr_network <- readRDS(url(paste0(zenodo_path,
      "lr_network_mouse_21122021.rds")))
weighted_networks <- readRDS(url(paste0(zenodo_path,
      "weighted_networks_nsga2r_final_mouse.rds")))
```

The ligand-target prior model is a matrix describing the potential that a ligand may regulate a target gene, and it is used to run the ligand activity analysis and target gene prediction. The ligand-receptor network contains information on potential ligand-receptor bindings, and it is used to identify potential ligands. Finally, the weighted ligand-receptor network contains weights representing the potential that a ligand will bind to a receptor, and it is used for visualization.

These networks were translated from human to mouse gene names using one-to-one orthologs when feasible, and one-to-many conversion was allowed when necessary (for instance, when one human gene symbol corresponded to two mouse gene symbols). Users that are interested in building prior models for other organisms can either create an organism-specific model using data sources relevant to that organism, or use the existing human NicheNet model to convert human gene symbols to their corresponding one-to-one orthologs in the organism of interest. However, this decision depends on one hand, the availability of data for the organism of interest and on the other, the homology between humans and the organism of interest. For instance, using the human model and converting gene symbols might work for primates, but creating a new model from species-specific data sources is better suited for organisms like Drosophila.

## Procedure

▲ **CRITICAL** The procedure described here also includes the sender-focused approach.

▲ **CRITICAL** As two conditions are present in this example dataset, the gene set of interest is chosen as the DE genes between these conditions in the receiver cell type, as shown in **Figure 2**.

▲ **CRITICAL** Properties of intermediate variables can be found in **Table 1**.

▲ **CRITICAL** Box 2 details the use of wrapper functions that can automatically run Steps 5-23.

**Feature extraction** ● **Timing** < 10 seconds

1. Load required libraries.

    ```r
    library(nichenetr)
    library(Seurat)
    library(tidyverse)
    ```

    **? TROUBLESHOOTING**

2. (Optional) For older Seurat objects, update it to be compatible with the currently installed Seurat version. For expression data with older gene symbols, convert them to more recent gene symbols.

    ```r
    seuratObj <- UpdateSeuratObject(seuratObj)
    seuratObj <- alias_to_symbol_seurat(seuratObj, "mouse")
    ```

3. Set the cell type annotation column as the identity of the Seurat object.

    ```r
    Idents(seuratObj) <- seuratObj$celltype
    ```

4. Define a "receiver" cell population.

    ```r
    receiver <- "CD8 T"
    ```

    ▲ **CRITICAL STEP** The receiver cell population can only consist of one cell type.
    **? TROUBLESHOOTING**

5. Determine which genes are expressed in the receiver cell population.

    ```r
    expressed_genes_receiver <- get_expressed_genes(receiver, seuratObj,
        pct = 0.05)
    ```

    ▲ **CRITICAL STEP** The function get_expressed_genes considers genes to be expressed if they have non-zero counts in a certain percentage of the cell population (by default set at 10%). Here, we have lowered the threshold to 5% (pct) as some of the ligands and receptors are very lowly expressed. Users are also free to define expressed genes differently in a way that fits their data.

6. Get a list of all receptors available in the ligand-receptor network, and define expressed receptors as genes that are in the ligand-receptor network and expressed in the receiver.

    ```r
    all_receptors <- unique(lr_network$to)
    expressed_receptors <- intersect(all_receptors, expressed_genes_receiver)
    ```

7. Define the potential ligands as all ligands whose cognate receptors are expressed.

```r
potential_ligands <- lr_network[lr_network$to %in% expressed_receptors, ]
potential_ligands <- unique(potential_ligands$from)
```

8. (Optional) For the sender-focused approach, define sender cell types and expressed genes in all populations combined. Then, filter potential ligands to those that are expressed in sender cells.

```r
sender_celltypes <- c("CD4 T", "Treg", "Mono", "NK", "B", "DC")
list_expressed_genes_sender <- lapply(sender_celltypes, function(celltype)
    {get_expressed_genes(celltype, seuratObj, pct = 0.05)})
expressed_genes_sender <- unique(unlist(list_expressed_genes_sender))
potential_ligands_focused <- intersect(
        potential_ligands, expressed_genes_sender)
```

9. Define the reference condition and condition of interest.

```r
condition_oi <- "LCMV"
condition_reference <- "SS"
```

▲ **CRITICAL STEP** The condition of interest is the condition after the CCC event has taken place, or the 'case' group in case-control studies. Here, it represents the condition after LCMV infection.

10. Define the gene set of interest that represents the cell-cell communication event to be studied. First, create a new Seurat object that only contains the receiver cell type. Then, perform DE analysis between the treatment conditions within the receiver cell type. Finally, define the gene set of interest as significantly DE genes, i.e., genes with adjusted *p*-value lower than or equal to 0.05 and absolute log-fold change greater than 0.25.

```r
seurat_obj_receiver <- subset(seuratObj, idents = receiver)
DE_table_receiver <- FindMarkers(object = seurat_obj_receiver,
    ident.1 = condition_oi, ident.2 = condition_reference,
    group.by = "aggregate",
    min.pct = 0.05)
geneset_oi <- DE_table_receiver[DE_table_receiver$p_val_adj <= 0.05 &
        abs(DE_table_receiver$avg_log2FC) >= 0.25, ]
geneset_oi <- rownames(geneset_oi)[rownames(geneset_oi) %in%
        rownames(ligand_target_matrix)]
```

▲ **CRITICAL STEP** By default, both genes that are up and downregulated are considered. Users can choose to focus on only one direction (typically upregulation) by removing the `abs()` function and adjusting the equality term to either `>= 0.25` or `<= -0.25` for up and downregulation, respectively. We recommend the gene set of interest to contain between 20 and 2000 genes for optimal ligand activity prediction. Moreover, the number of background genes should ideally be at least 10 times greater than those of the gene set of interest.

11. Determine background genes as all the genes expressed in the receiver cell type that are also in the ligand-target matrix.

```r
background_expressed_genes <- expressed_genes_receiver[
    expressed_genes_receiver %in% rownames(ligand_target_matrix)]
```

Ligand activity analysis and downstream prediction ● **Timing** < 10 seconds

12 Perform the ligand activity analysis, then sort the ligands based on the area under the precision-recall curve.

```
ligand_activities <- predict_ligand_activities(
    geneset = geneset_oi,
    background_expressed_genes = background_expressed_genes,
    ligand_target_matrix = ligand_target_matrix,
    potential_ligands = potential_ligands)
ligand_activities <- ligand_activities[order(ligand_activities$aupr_corrected,
    decreasing = TRUE), ]
```

**? TROUBLESHOOTING**

13 (Optional) If performing the sender-focused approach, subset the ligand activities to only contain expressed ligands.

```
ligand_activities_all <- ligand_activities
ligand_activities <- ligand_activities[
    ligand_activities$test_ligand %in% potential_ligands_focused, ]
```

▲ **CRITICAL STEP** When using the sender-agnostic approach, simply replace `ligand_activities` with `ligand_activities_all` in Steps 14 and 20.

14 Obtain the names of the top 30 ligands.

```
best_upstream_ligands <- top_n(ligand_activities, 30,
    aupr_corrected)$test_ligand
```

▲ **CRITICAL STEP** Box 3 describes a method for assessing the quality of predicted ligands.

15 Infer which genes in the gene set of interest have the highest regulatory potential for each top-ranked ligand.

```
active_ligand_target_links_df <- lapply(best_upstream_ligands,
    get_weighted_ligand_target_links,
    geneset = geneset_oi,
    ligand_target_matrix = ligand_target_matrix,
    n = 200)
active_ligand_target_links_df <-
    drop_na(bind_rows(active_ligand_target_links_df))
```

▲ **CRITICAL STEP** The function `get_weighted_ligand_target_links` will return genes that are in the gene set of interest and are the top n targets of a ligand (default: n = 200).

16 Similarly, identify which receptors have the highest interaction potential with the top-ranked ligands.

```
ligand_receptor_links_df <- get_weighted_ligand_receptor_links(
    best_upstream_ligands, expressed_receptors,
    lr_network, weighted_networks$lr_sig)
```

Visualizations  ● **Timing** < 10 seconds

▲ **CRITICAL**    Visualizations covered in this section are shown in **Figure 3** and include: heatmaps of ligand-target regulatory potential (Steps 17-18), ligand-receptor interaction potential (Step 19), ligand activity (Step 20), and log-fold change of ligands between treatment conditions (Steps 21-22); a dot plot of cell type expression and percentage (Step 23); a line plot comparing ligand rankings between the sender-agnostic and -focused approach (Step 24); chord diagrams (Steps 25-29); and a signaling graph (Steps 30-31).

▲ **CRITICAL**    This section can be followed in its entirety only for the sender-focused approach (i.e., if all optional code in the previous sections have been executed); for the sender-agnostic approach, only Steps 17-20 and Steps 30-31 are relevant.

▲ **CRITICAL**    The code from this section will produce slightly different visualizations from **Figure 3** due to adjustments in `theme` and `cutoff` values; the exact code for reproducing the figure is available in our GitHub repository.

17  Prepare the weighted ligand-target data frame for visualization by transforming it into matrix.

```
active_ligand_target_links <- prepare_ligand_target_visualization(
    ligand_target_df = active_ligand_target_links_df,
    ligand_target_matrix = ligand_target_matrix,
    cutoff = 0.25)
```

▲ **CRITICAL STEP**    By default, regulatory potentials lower than the 25$^{th}$ percentile are set to zero for visualization clarity. This `cutoff` parameter can freely be tuned by the user.

18  Order the rows to follow the rankings of the ligands, and the columns alphabetically. Create the heatmap (**Figure 3a**).

```
order_ligands <- rev(intersect(best_upstream_ligands,
    colnames(active_ligand_target_links)))
order_targets <- intersect(unique(active_ligand_target_links_df$target),
    rownames(active_ligand_target_links))
vis_ligand_target <- t(active_ligand_target_links[order_targets,order_ligands])

(make_heatmap_ggplot(vis_ligand_target,
    y_name = "Prioritized ligands", x_name = "Predicted target genes",
    color = "purple", legend_title = "Regulatory potential") +
    scale_fill_gradient2(low = "whitesmoke", high = "purple"))
```

**? TROUBLESHOOTING**

19  Create a heatmap for ligand-receptor interactions (**Figure 3b**).

```
vis_ligand_receptor_network <- prepare_ligand_receptor_visualization(
    ligand_receptor_links_df, best_upstream_ligands,
    order_hclust = "receptors")
(make_heatmap_ggplot(t(vis_ligand_receptor_network),
    y_name = "Ligands", x_name = "Receptors",
    color = "mediumvioletred", legend_title = "Prior interaction potential"))
```

20  Create a heatmap of the ligand activity measure (**Figure 3c**).

```
ligand_aupr_matrix <- column_to_rownames(ligand_activities, "test_ligand")
```

```r
ligand_aupr_matrix <- ligand_aupr_matrix[rev(best_upstream_ligands),
    "aupr_corrected", drop=FALSE]
vis_ligand_aupr <- as.matrix(ligand_aupr_matrix, ncol = 1)
(make_heatmap_ggplot(vis_ligand_aupr,
    "Prioritized ligands", "Ligand activity",
    legend_title = "AUPR", color = "darkorange") +
    theme(axis.text.x.top = element_blank()))
```

21  For each cell type, compute the log-fold change of the top-ranked ligands between treatment conditions.

```r
celltype_order <- levels(Idents(seuratObj))
DE_table_top_ligands <- lapply(
    celltype_order[celltype_order %in% sender_celltypes],
    get_lfc_celltype,
    seurat_obj = seuratObj, condition_colname = "aggregate",
    condition_oi = condition_oi, condition_reference = condition_reference,
    celltype_col = "celltype",
    min.pct = 0, logfc.threshold = 0,
    features = best_upstream_ligands
)
DE_table_top_ligands <- reduce(DE_table_top_ligands, full_join)
DE_table_top_ligands <- column_to_rownames(DE_table_top_ligands, "gene")
```

22  Create the heatmap (**Figure 3d**).

```r
vis_ligand_lfc <- as.matrix(DE_table_top_ligands[rev(best_upstream_ligands), ])
(make_threecolor_heatmap_ggplot(vis_ligand_lfc,
    "Prioritized ligands", "LFC in Sender",
    low_color = "midnightblue", mid_color = "white",
    mid = median(vis_ligand_lfc), high_color = "red",
    legend_title = "LFC"))
```

**? TROUBLESHOOTING**

23  Create a dot plot showing the average expression of ligands per cell type, as well as the percentage of cells from the cell type expressing the ligands (**Figure 3e**).

```r
DotPlot(subset(seuratObj, celltype %in% sender_celltypes),
    features = rev(best_upstream_ligands), cols = "RdYlBu") +
    coord_flip() +
    scale_y_discrete(position = "right")
```

**? TROUBLESHOOTING**

24  Create a line plot comparing the rankings between the sender-agnostic and sender-focused approach (**Figure 3f**).

```r
(make_line_plot(ligand_activities = ligand_activities_all,
    potential_ligands = potential_ligands_focused))
```

25. To create a ligand-target chord diagram, assign each ligand to a specific cell type.

    ```
    ligand_type_indication_df <- assign_ligands_to_celltype(seuratObj,
        best_upstream_ligands[1:20], celltype_col = "celltype")
    ```

    ▲ CRITICAL STEP   A ligand is only assigned to a cell type if that cell type is the only one to show an average expression of that ligand that is higher than the mean + one standard deviation across all cell types. Otherwise, it is assigned to "General".

26. Using the weighted ligand-target data frame from Step 15, group target genes and filter out the lowest 40% of the regulatory potentials.

    ```
    active_ligand_target_links_df$target_type <- "LCMV-DE"
    circos_links <- get_ligand_target_links_oi(ligand_type_indication_df,
        active_ligand_target_links_df, cutoff = 0.40)
    ```

    ▲ CRITICAL STEP   In this case, there is only one grouping of target genes (DE genes after LCMV infection), but users can define multiple target gene groups if applicable. In case the resulting chord diagram is still overcrowded, users may adjust the `cutoff` parameter to filter out even more ligand-target links.

27. Assign colors to cell types and target gene groups. Then, prepare the data frame for visualization: the function assigns colors to ligands and targets and calculates gaps between sectors of the chord diagram.

    ```
    ligand_colors <- c("General" = "#377EB8", "NK" = "#4DAF4A", "B" = "#984EA3",
        "Mono" = "#FF7F00", "DC" = "#FFFF33", "Treg" = "#F781BF",
        "CD8 T"= "#E41A1C")
    target_colors <- c("LCMV-DE" = "#999999")
    vis_circos_obj <- prepare_circos_visualization(circos_links,
        ligand_colors = ligand_colors, target_colors = target_colors)
    ```

28. Draw the chord diagram (**Figure 3g**).

    ```
    make_circos_plot(vis_circos_obj, transparency = TRUE)
    ```

    ▲ CRITICAL STEP   The degree of transparency is determined by the regulatory potential value of a ligand-target interaction. If `transparency = FALSE`, all links will be opaque.

29. To create a ligand-receptor chord diagram, perform Steps 26-28 using the weighted ligand-receptor data frame from Step 16.

    ```
    lr_network_top_df <- rename(ligand_receptor_links_df, ligand=from, target=to)
    lr_network_top_df$target_type = "LCMV_CD8T_receptor"
    lr_network_top_df <- inner_join(lr_network_top_df, ligand_type_indication_df)
    receptor_colors <- c("LCMV_CD8T_receptor" = "#E41A1C")
    vis_circos_receptor_obj <- prepare_circos_visualization(lr_network_top_df,
        ligand_colors = ligand_colors, target_colors = receptor_colors)
    make_circos_plot(vis_circos_receptor_obj, transparency = TRUE,
        link.visible = TRUE)
    ```

▲ **CRITICAL STEP** As prepare_circos_visualization accesses "target" and "target_type" columns, it is necessary to rename the columns accordingly even though the data frame contains receptor and not target gene information. When drawing the plot, the argument link.visible = TRUE is also necessary for making all links visible, since no cutoff is used to filter out ligand-receptor interactions.

30  To create a signaling graph, first download the ligand-transcription factor matrix. Then, extract the most highly weighted paths from the ligand to the target genes of interest.

```
ligand_tf_matrix <- readRDS(url("https://zenodo.org/record/7074291/
    files/ligand_tf_matrix_nsga2r_final_mouse.rds"))
ligands_oi <- "Ebi3"
targets_oi <- c("Irf1", "Irf9")
active_signaling_network <- get_ligand_signaling_path(
    ligands_all = ligands_oi, targets_all = targets_oi,
    weighted_networks = weighted_networks,
    ligand_tf_matrix = ligand_tf_matrix,
    top_n_regulators = 4, minmax_scaling = TRUE)
```

▲ **CRITICAL STEP** The number of regulators that are extracted can be adjusted by using top_n_regulators. By setting minmax_scaling = TRUE, we perform min-max scaling to make the weights between the signaling and gene regulatory network more comparable. Additionally, it is possible to check which data sources support the inferred pathway by using the function infer_supporting_datasources. This would require separate signaling and gene regulatory networks as input (see Box 1 for the code to download these networks).

31  Convert the data frames into a DiagrammeR object, and render the signaling graph (**Figure 3h**).

```
signaling_graph <- diagrammer_format_signaling_graph(
    signaling_graph_list = active_signaling_network,
    ligands_all = ligands_oi, targets_all = targets_oi,
    sig_color = "indianred", gr_color = "steelblue")
DiagrammeR::render_graph(signaling_graph, layout = "tree")
```

Prioritization of ligand-receptor pairs     ● **Timing** 10 seconds

▲ **CRITICAL** This section is only applicable for the sender-focused approach.

▲ **CRITICAL** Whereas Steps 12-14 only prioritize ligands based on ligand activity, this section incorporates relative expression and differential expression to further prioritize ligand-receptor pairs associated with specific sender and receiver cell types.

32  Filter the ligand-receptor network to only contain expressed interactions.

```
lr_network_filtered <- filter(lr_network,
    from %in% potential_ligands_focused &
    to %in% expressed_receptors)[, c("from", "to")]
```

33  Calculate the values required for prioritization, including DE between cell types, average expression, and DE between conditions. To use the wrapper function, follow option A. To calculate these values step-by-step, follow option B.

(A) **Using the wrapper function.**
(i) Run the wrapper function.

```
info_tables <- generate_info_tables(
    seuratObj,
    celltype_colname = "celltype",
    senders_oi = sender_celltypes,
    receivers_oi = receiver,
    lr_network = lr_network_filtered,
    condition_colname = "aggregate",
    condition_oi = condition_oi,
    condition_reference = condition_reference,
    scenario = "case_control")
```

▲ **CRITICAL STEP** In case of steady-state datasets, use `scenario="one_condition"` and do not provide anything to the `condition_` arguments.

(ii) Assign the output of the wrapper function to variables.

```
processed_DE_table <- info_tables$sender_receiver_de
processed_expr_table <- info_tables$sender_receiver_info
processed_condition_markers <- info_tables$lr_condition_de
```

(B) **Running step-by-step calculations.**
(i) Calculate DE between cell types within the condition of interest.

```
DE_table <- FindAllMarkers(subset(seuratObj, subset = aggregate == "LCMV"),
    min.pct = 0, logfc.threshold = 0, return.thresh = 1,
    features = unique(unlist(lr_network_filtered)))
```

▲ **CRITICAL STEP** The `min.pct`, `logfc.threshold`, and `return.thresh` parameters are required for the function to return p-values for all features and cell types.

(ii) Calculate average expression of each gene per cell type.

```
expression_info <- get_exprs_avg(seuratObj, "celltype",
    condition_colname = "aggregate", condition_oi = condition_oi,
    features = unique(unlist(lr_network_filtered)))
```

(iii) Calculate DE between conditions.

```
condition_markers <- FindMarkers(seuratObj,
    ident.1 = condition_oi, ident.2 = condition_reference,
    group.by = "aggregate",
    min.pct = 0, logfc.threshold = 0,
    features = unique(unlist(lr_network_filtered)))
condition_markers <- rownames_to_column(condition_markers, "gene")
```

(iv) Process the data frames from (i)-(iii) to follow the same format.

```
processed_DE_table <- process_table_to_ic(
    DE_table,
    table_type = "celltype_DE",
    lr_network_filtered,
    senders_oi = sender_celltypes, receivers_oi = receiver)
```

```r
    processed_expr_table <- process_table_to_ic(
        expression_info,
        table_type = "expression",
        lr_network_filtered)
    processed_condition_markers <- process_table_to_ic(
        condition_markers,
        table_type = "group_DE",
        lr_network_filtered)
```

▲ **CRITICAL STEP**   This function converts the input into a table containing sender-ligand-receptor-receiver columns (for cell type DE and expression values) or ligand-receptor columns (for condition DE values).

34  Generate the prioritization table containing rankings of cell-type-specific, ligand-receptor interactions.

```r
    prioritized_table <- generate_prioritization_tables(
        processed_expr_table,
        processed_DE_table,
        ligand_activities,
        processed_condition_markers,
        scenario = "case_control")
```

▲ **CRITICAL STEP**   The "case_control" scenario sets all weights to one, while the "one_condition" scenario sets condition specificity to zero and the remaining weights to one. Users may provide custom weights in a named vector to the argument `prioritizing_weights`. The vector must contain the following names, which correspond to the following criteria:
- `activity_scaled`: ligand activity of the ligand
- `de_ligand`: upregulation of the ligand in a sender cell type compared to other cell types
- `de_receptor`: upregulation of the receptor in a receiver cell type
- `exprs_ligand`: average expression of the ligand in the sender cell type
- `exprs_receptor`: average expression of the receptor in the receiver cell type
- `ligand_condition_specificity`: condition-specificity of the ligand across all cell types
- `receptor_condition_specificity`: condition-specificity of the receptor across all cell types

35  Create a mushroom plot depicting ligand expression on one semicircle, and receptor expression on the other (**Figure 3i**).

```r
    make_mushroom_plot(prioritized_table, top_n = 30,
        show_all_datapoints = TRUE, true_color_range = TRUE, show_rankings = TRUE)
```

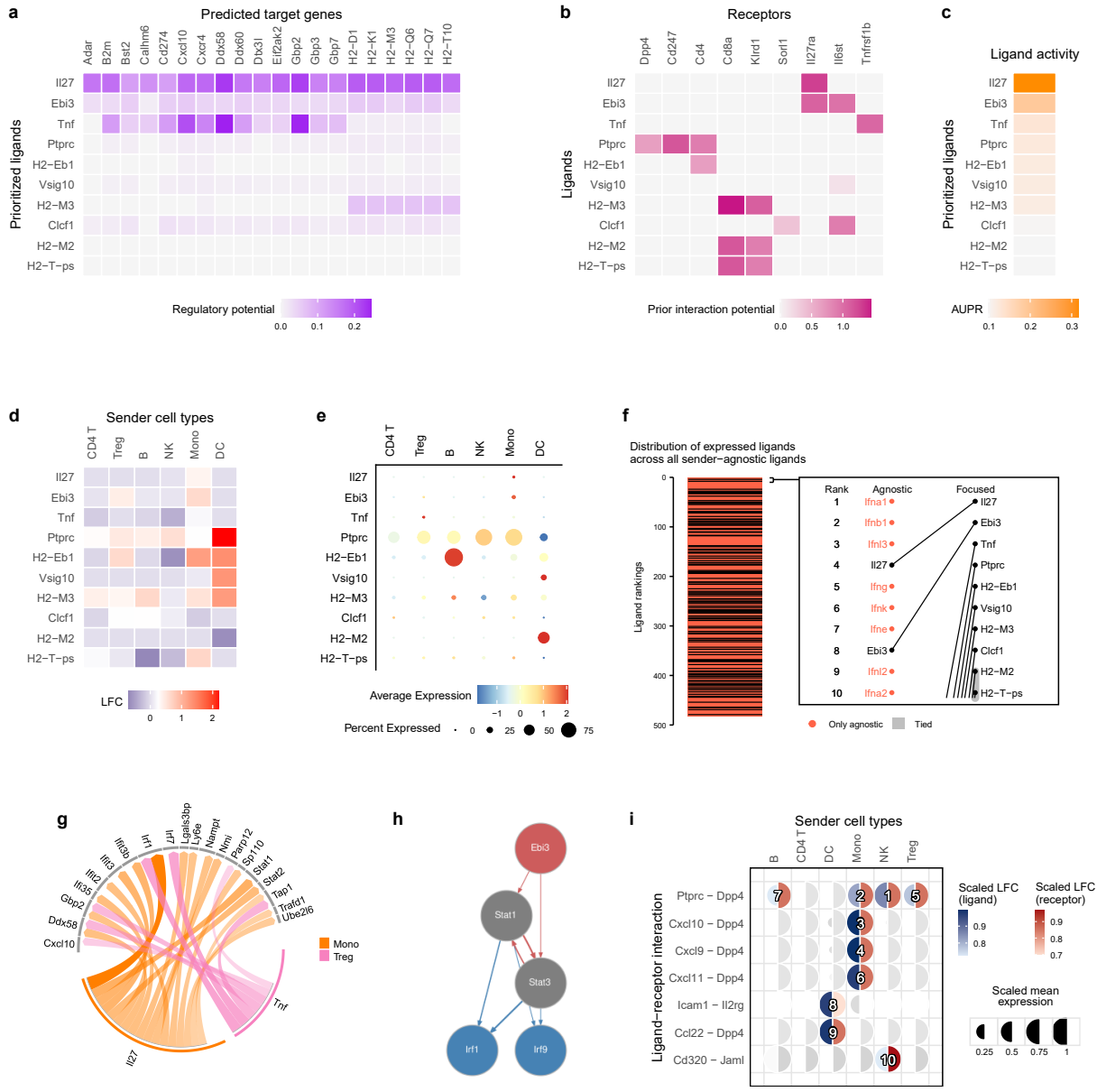

**Figure 3. Example visualizations from the sender-focused approach.** For visibility, only the top 10 ligands are shown with minor adjustments to the plot style elements. The code to reproduce this figure can be found in the paper's repository. Note that when using the sender-agnostic approach, only plots **a-c** and **h** can be created. **(a)** A heatmap displaying the target gene candidates from the gene set of interest. **(b)** Receptors with the highest interaction potential with the prioritized ligands are depicted and hierarchically clustered. **(c)** The ligand activity (i.e., area under the precision-recall curve) measures how well each ligand predicts the target gene set. **(d)** The log-fold change (LFC) of each ligand pre- and post-LCMV treatment across various cell types. This heatmap aids in identifying condition-specific upregulation, such as Il27 in monocytes and Ptprc in dendritic cells. **(e)** The percentage of cells per cell type expressing each ligand on average, used as a complement to **(d)**. **(f)** The distribution of expressed ligands in the sender-focused approach across all ligands in the sender-agnostic approach (left), and the shift in rankings between the two approaches (inset). Ligands in red are only present in the agnostic approach. **(g)** A chord diagram linking cell-type-specific ligands with their targets. To enhance visibility, ligand-target interactions below the 95% quartile are omitted. **(h)** A possible signaling path from Ebi3 to Irf1 and Irf9. Only one regulatory for each target gene is shown. **(i)** A mushroom plot depicting the rankings of cell type-specific ligand-receptor pairs, with CD8 T cells as the receiver. Each semicircle encodes the scaled LFC and mean expression for either the ligand or receptor. Unlike **(d)** which measures the LFC between conditions, the scaled LFC here is calculated across all cell types within the condition of interest.

**Table 1. Properties of objects and intermediate variables.**

| Step | Variable | Dimensions | Type | Preview (row and column names in bold) |
|---|---|---|---|---|
| Material | *ligand_target_matrix* | 22521 × 1287 | Matrix |                  **2300002M23Rik**   **2610528A11Rik**<br>**0610005C13Rik**  0.000000e+00    0.000000e+00<br>**0610009B22Rik**  0.000000e+00    0.000000e+00<br>**0610009L18Rik**  8.872902e-05    4.977197e-05 |
| Material | *lr_network* | 5668 × 4 | Data frame | **from**          **to**     **database**  **source**<br>2300002M23Rik  Ddr1   omnipath   omnipath<br>2610528A11Rik  Gpr15  omnipath   omnipath<br>9530003J23Rik  Itgal  omnipath   omnipath |
| Material | *weighted_networks$ lr_sig* | 3865137 × 3 | Data frame | **from**          **to**    **weight**<br>0610010F05Rik  App    0.110<br>0610010F05Rik  Cat    0.0673<br>0610010F05Rik  H1f2   0.0660 |
| 5 | *expressed_genes_ receiver* | 3903 | Vector | "Atraid"       "Tmem248"     "Tsr3"       "Sf3b6"<br>"0610010K14Rik" "0610012G03Rik" |
| 6 | *expressed_receptors* | 113 | Vector | "Itgal"      "Notch1"    "Tspan14"   "Itga4"<br>"Itgb1"      "Il6ra" |
| 7 | *potential_ligands* | 483 | Vector | "9530003J23Rik" "Adam10"     "Adam11"    "Adam12"<br>"Adam15"     "Adam17" |
| 8 | *expressed_genes_ sender* | 8492 | Vector | "Atraid"       "Tmem248"     "Sf3b6"       "0610010K14Rik"<br>"0610012G03Rik" "Sptssa" |
| 8 | *potential_ligands_ focused* | 127 | Vector | "Adam10"    "Adam15"    "Adam17"    "Adam9"<br>"Adgre5"    "Alcam" |
| 10 | *geneset_oi* | 260 | Vector | "Ifi27l2b"   "Irf7"      "Ly6a"      "Stat1"<br>"Ly6c2"     "Ifit3" |
| 11 | *background_ expressed_genes* | 3476 | Vector | "Atraid"       "Tmem248"     "Tsr3"       "Sf3b6"<br>"0610010K14Rik" "0610012G03Rik" |
| 12 | *ligand_activities* (In Step 13, renamed to *ligand_activities_all*) | 483 × 5 | Data frame | **test_ligand**  **auroc**  **aupr**  **aupr_corrected**  **pearson**<br>Ifna1       0.714  0.433  0.358           0.498<br>Ifnb1       0.711  0.401  0.327           0.433<br>Ifnl3       0.683  0.392  0.317           0.433 |
| 13 | *ligand_activities* | 127 × 5 | Data frame | **test_ligand**  **auroc**  **aupr**  **aupr_corrected**  **pearson**<br>Il27        0.682  0.391  0.316           0.445<br>Ebi3        0.666  0.264  0.189           0.256<br>Tnf         0.671  0.205  0.131           0.249 |
| 14 | *best_upstream_ligands* | 30 | Vector | "Il27"       "Ebi3"      "Tnf"       "Ptprc"<br>"H2-Eb1"     "Vsig10" |
| 15 | *active_ligand_target_ links_df* (for sender-focused approach) | 656 × 3 | Data frame | **ligand**  **target**  **weight**<br>Il27    Adar    0.163<br>Il27    B2m     0.170<br>Il27    Bst2    0.111 |
| 33 | *processed_DE_table* | 1272 × 13 | Data frame | **sender**  **receiver**  **ligand**  **receptor**  **lfc_ligand**  …<br>DC      CD8 T     Ccl5     Cxcr3     6.432<br>Mono    CD8 T     Lyz2     Itgal     5.493<br>DC      CD8 T     H2-M2    Cd8a     3.416 |
| 33 | *processed_expr_table* | 10535 × 7 | Data frame | **sender**  **receiver**  **ligand**  **receptor**  **avg_ligand**  …<br>DC      Mono      B2m      Tap1      216.<br>DC      NK        B2m      Klrd1     216.<br>DC      B         B2m      Tap1      216. |
| 33 | *processed_condition_ markers* | 215 × 9 | Data frame | **ligand**  **receptor**  **lfc_ligand**  **lfc_receptor**  …<br>H2-Ab1   Cd4      2.402       0.116<br>Cxcl10   Dpp4     1.607       0.352<br>B2m      Tap1     0.707       1.139 |
| 34 | *prioritized_table* | 1272 × 51 | Data frame | **sender**  **receiver**  **ligand**  **receptor**  **lfc_ligand**  …<br>NK      CD8 T     Ptprc    Dpp4      0.596<br>Mono    CD8 T     Ptprc    Dpp4      0.438<br>Mono    CD8 T     Cxcl10   Dpp4      4.27 |

**Box 1.** Constructing your own prior model

As the NicheNet prior model was constructed by integrating ligand-receptor, signaling, and gene regulatory databases, it is possible to replace each of these networks with external data sources. Constructing a customized prior model thus requires three directed networks, represented as data frames comprising three columns: "from", "to", and "source". The "source" column contains the originating database of the interaction. As the reliability of each database can vary, we optimized the weights of each data source based on our validation procedure (as explained in Comparison with other methods). These optimized weights (optimized_source_weights_df), along with hyperparameters for model construction (hyperparameter_list), are provided in the NicheNet package. Key hyperparameters include correction factors for dominant hubs in the ligand-signaling and gene regulatory networks, as well as the central damping factor of the Personalized PageRank algorithm, the network propagation mechanism used to determine the ligand-target regulatory potential.

```r
# Download each part of the network
zenodo_path <- "https://zenodo.org/record/7074291/files/"
lr_network <- readRDS(url(paste0(zenodo_path, "lr_network_human_21122021.rds")))
sig_network <- readRDS(url(paste0(zenodo_path,
    "signaling_network_human_21122021.rds")))
gr_network <- readRDS(url(paste0(zenodo_path, "gr_network_human_21122021.rds")))

# Aggregate the individual data sources in a weighted manner to obtain
# a weighted integrated signaling network
weighted_networks <- construct_weighted_networks(
    lr_network = lr_network,
    sig_network = sig_network,
    gr_network = gr_network,
    source_weights_df = rename(optimized_source_weights_df,
                                weight = avg_weight))

# Downweigh the importance of signaling and gene regulatory hubs
# Use the optimized parameters of this
weighted_networks <- apply_hub_corrections(
    weighted_networks = weighted_networks,
    lr_sig_hub = hyperparameter_list[
    hyperparameter_list$parameter == "lr_sig_hub",]$avg_weight,
    gr_hub = hyperparameter_list[
    hyperparameter_list$parameter == "gr_hub",]$avg_weight)

# In this example, we will calculate target gene regulatory potential scores for
# TNF and the combination TNF+IL6
# To compute it for all 1248 ligands (~1 min):
# ligands <- as.list(unique(lr_network$from))
ligands <- list("TNF", c("TNF","IL6"))
ligand_target_matrix = construct_ligand_target_matrix(
    weighted_networks = weighted_networks,
    ligands = ligands,
    algorithm = "PPR",
    damping_factor = hyperparameter_list[
    hyperparameter_list$parameter == "damping_factor",]$avg_weight,
    ltf_cutoff = hyperparameter_list[
    hyperparameter_list$parameter == "ltf_cutoff",]$avg_weight)
```

A frequent use case is one where users are interested in replacing the ligand-receptor network in NicheNet with those of expression permutation tools in order to make their results more comparable. To achieve this, we recommend employing LIANA[30], a comprehensive CCC framework that integrates both resources and computational algorithms for ligand-receptor interaction inference. The `show_resources` function is used to check which resources are present in LIANA, and `select_resource` returns a data frame of the interactions in that resource. The `decomplexify` function of LIANA is necessary for this integration, as it separate receptors into their respective subunits. Note that unlike before, `source_weights_df` only represents unoptimized weights, where the weight of every data source is 1.

```r
devtools::install_github("saezlab/liana")
liana_db <- liana::decomplexify(liana::select_resource("CellPhoneDB")[[1]])
liana_db <- rename(liana_db, from = source_genesymbol, to = target_genesymbol)
liana_db$source <- "liana"
liana_db <- select(liana_db, from, to, source)

# Change source weights data frame (but in this case all source weights are 1)
source_weights <- add_row(source_weights_df, source = "liana",
       weight = 1, .before = 1)

# Construct weighted network as before
weighted_networks <- construct_weighted_networks(
       lr_network = liana_db,
       sig_network = sig_network,
       gr_network = gr_network,
       source_weights_df = source_weights)
```

**Box 2.** Wrapper functions

To streamline the NicheNet analysis, we introduce three wrapper functions that automate Steps 5-23. The function `nichenet_seuratobj_aggregate` calculates the gene set of interest as the DE genes between two conditions within the receiver cell type. This function can be used to replicate the analysis in this paper as follows:

```r
nichenet_output <- nichenet_seuratobj_aggregate(
    seurat_obj = seuratObj,
    receiver = "CD8 T",
    sender = c("CD4 T","Treg", "Mono", "NK", "B", "DC"),
    condition_colname = "aggregate",
    condition_oi = "LCMV", condition_reference = "SS",
    ligand_target_matrix = ligand_target_matrix,
    lr_network = lr_network,
    weighted_networks = weighted_networks)
```

Additionally, the sender-agnostic approach can be explicitly run by setting `sender = "undefined"`.

The resulting object is a list comprising various components, including the gene set of interest and background genes that were used for the analysis (`geneset_oi` and `background_expressed_genes`), output from the ligand activity analysis (`ligand_activities`), targets and receptors corresponding to the identified ligands (`top_targets` and `top_receptors`), and visualizations (`ligand_target_heatmap`, `ligand_expression_dotplot`, `ligand_receptor_heatmap`, etc.).

Another wrapper function, `nichenet_seuratobj_cluster_de`, calculates DE genes between two cell types as the gene set of interest. Additionally, when filtering for potential ligands, we only consider expressed ligands whose receptors are expressed by the "reference" receiver cell type. This function should only be used for specific biological scenarios, as shown in **Figure 2**. An example of using this function for the scenario where cell type differentiation occurs due to its niche is as follows:

```r
nichenet_seuratobj_cluster_de(
    seurat_obj = seurat_obj,
    receiver_affected = differentiated_celltype,
    receiver_reference = progenitor_cell,
    sender = niche_celltypes,
    ligand_target_matrix, lr_network, weighted_networks)
```

The final wrapper function, `nichenet_seuratobj_aggregate_cluster_de`, combines the aforementioned wrappers. The gene set of interest is calculated as the DE genes between the affected receiver cell type under the condition of interest and the reference receiver cell type in the reference condition. This could be used for cases where only a subset of the affected and reference cell types should be compared, e.g., from different time points:

```r
nichenet_seuratobj_aggregate_cluster_de(
    seurat_obj = seurat_obj,
    receiver_affected = differentiated_celltype,
    receiver_reference = progenitor_cell,
    condition_colname = "timepoint",
    condition_oi = "t1", condition_reference = "t0",
    sender = niche_celltypes,
    ligand_target_matrix, lr_network, weighted_networks)
```

**Box 3.** Assessing the quality of predicted ligands

We can assess the quality of prioritized ligands by constructing a random forest model built using the top 30 predicted ligands. The aim is to evaluate its ability to predict if a particular gene belongs to the target gene set. Using k-fold cross-validation, 1/k of the target gene set is isolated as "unseen" data. The performance of the model can then be evaluated using classification metrics like AUPR and AUROC, or using Fisher's exact test on the confusion matrix.

```r
# Define cross-validation folds and number of iterations
k <- 3; n <- 10

# Build random forest model and obtain prediction values
predictions_list <- lapply(1:n, assess_rf_class_probabilities,
       folds = k, geneset = geneset_oi,
       background_expressed_genes = background_expressed_genes,
       ligands_oi = best_upstream_ligands,
       ligand_target_matrix = ligand_target_matrix)

# Get classification metrics of the models, then calculate mean across all rounds
performances_cv <- bind_rows(lapply(predictions_list,
       classification_evaluation_continuous_pred_wrapper))
colMeans(performances_cv)

# Calculate fraction of target genes and non-target genes
# that are among the top 5% predicted targets
fraction_cv <- bind_rows(lapply(predictions_list,
       calculate_fraction_top_predicted, quantile_cutoff = 0.95), .id = "round")

# In this case, ~30% of target genes are in the top targets
# compared to ~1% of the non-target genes
mean(filter(fraction_cv, true_target)$fraction_positive_predicted)
mean(filter(fraction_cv, !true_target)$fraction_positive_predicted)

# Perform Fischer's exact test
lapply(predictions_list, calculate_fraction_top_predicted_fisher,
       quantile_cutoff = 0.95)

# Get which genes had the highest prediction values
top_predicted_genes <- lapply(1:n, get_top_predicted_genes, predictions_list)
top_predicted_genes <- reduce(top_predicted_genes, full_join,
       by = c("gene","true_target"))
```

# Troubleshooting

| Step | Problem | Possible reason(s) | Solution(s) |
|---|---|---|---|
| Input | I am unable to store my data as a Seurat object | (A) Bulk expression data (B) Analysis pipeline is based on another object format | Use analogous functions (if any) from the library of interest to perform the needed task, e.g., for differential expression testing: (A) `exactTest` in edgeR; `DESeq` in DESeq2 (B) `scran::scoreMarkers` for SingleCellExperiment objects |
| 4 | There are multiple receiver cell types | Part of the biological hypothesis | Run the pipeline multiple times for each cell type separately |
| 12 | AUPR values are near zero or negative | (A) The gene set of interest is not well defined (B) The background gene set is too small | (A) Change parameters of `FindMarkers`, or find an alternative way to define the gene set of interest (B) Change the definition of expressed genes, or use all genes from the ligand-target matrix as background |
| 18 | Not all genes of the gene set of interest are in the final ligand-target heatmap | The heatmap only shows genes that are top predicted targets of at least one of the top ligands | In `get_weighted_ligand_target_links`, increase `n` to consider more top targets of a ligand |
| 22 | Some top ligands that are supposed to be inducing in the condition of interest are higher expressed in control than condition of interest | The ligand activity analysis does not take expression data into account | (A) After the ligand activity analysis, filter for ligands that are upregulated in the condition of interest (recommended) (B) Before the ligand activity analysis, perform a DE analysis on the sender cells and only consider upregulated ligands as potential ligands (C) Use the downstream prioritization scheme, which takes upregulation of ligands into account |
| 23 | Lowly expressed ligands are in the top ligands | The ligand activity analysis does not take expression data into account | Use the prioritization scheme |

# Timing

The timing information below only lists the time required to execute the code. In practice, users will need to make decisions about the workflow (**Figure 2**) and inspect the results, which may also take a considerable amount of time. The execution time was measured using the example dataset with 13,541 genes and 5,027 cells. For larger datasets, longer runtimes may be observed for DE analyses, e.g., those performed during target gene set selection (Step 10), visualization (Step 21) and prioritization (Step 33). However, the ligand activity analysis and target gene prediction can scale to any dataset size as they only make use of the three extracted features. The code was run on a computer with an Intel i7-11850H processor operating at 2.5GHz and 64GB of RAM, running CentOS version 8.

Steps 1-11, feature extraction: < 10 seconds
Steps 12-16, ligand activity analysis and downstream prediction: < 10 seconds
Steps 17-31, visualization: < 10 seconds
Steps 32-35, prioritization: ~ 10 seconds
Box 1, full model construction with all ligands: ~1 minute
Box 2, wrapper function: < 10 seconds
Box 3, quality assessment: ~1 minute (scales linearly with number of iterations)

## Anticipated results

For the ligand activity analysis, NicheNet generates a detailed data frame containing scores that assess how well each ligand explains observed changes in gene expression using the relevant gene set (**Table 1**, Steps 12-13). Four scoring metrics are incorporated: Pearson correlation, area under the ROC curve, and both unadjusted and adjusted AUPR. Based on our validation, NicheNet v2 identified the (adjusted) AUPR as the best measure of ligand activity. As a result, the adjusted AUPR is set as the default measure for users. By using these ligand rankings, users can then infer receptors and target genes that are most likely to be involved in the CCC process. Additionally, users have the option to compute the rankings of cell-type-specific ligand-receptor interactions (**Table 1**, Step 34). The resulting data frame also provides the underlying data used for these rankings, including ligand/receptor differential expression and average expression between cell types, and differential expression between conditions.

## Data availability

The original NICHE-seq data can be accessed at the Gene Expression Omnibus with accession number GSE104054. The processed Seurat object can be downloaded on Zenodo (https://zenodo.org/record/3531889).

## Code availability

NicheNet is publicly available on https://github.com/saeyslab/nichenetr/ as an R package. The code shown in this paper and the code for reproducing the figures can be found on https://github.com/saeyslab/nichenet_protocol.

## Acknowledgements

C.S. is funded by the Ghent University Special Research Fund [grant number BOF21-DOC-105], R.S. is funded by the Flemish Government under the Flanders AI Research Program, and Y.S is funded by Ghent University Special Research Fund [grant number BOF18-GOA-024], the Belgian Excellence of Science (EOS) program, FWO SBO [grant number S001121N] and the Leducq project "Cellular and Molecular Drivers of Acute Aortic Dissections".

## Author contributions

R.B. conceptualized the NicheNet algorithm. R.B. and C.S. wrote the code. R.S. and Y.S. supervised the work. C.S. wrote the manuscript. All authors edited, read and approved the manuscript.